\documentclass[twocolumn,showpacs,prb,aps]{revtex4}

\usepackage{graphicx}
\usepackage{bm}

\newcommand{\new}{\newcommand}
\new{\da}{\dagger}
\new{\up}{\uparrow}
\new{\down}{\downarrow}
\new{\nyo}{\tilde}
\new{\sig}{\sigma}
\new{\bold}{\boldmath}
\new{\ra}{\rangle}
\new{\la}{\langle}
\new{\nearl}{\stackrel{\textstyle <}{\sim}}
\new{\nearg}{\stackrel{\textstyle >}{\sim}}
\new{\nc}{\nyo{c}}
\new{\nn}{\nyo{n}}
\new{\ri}{\right}
\new{\lef}{\left}
\new{\barm}{\bar{m}}

\begin{document}


\title{Ferromagnetism in the one-dimensional Hubbard model
with orbital degeneracy: \\
From low to high electron density}

\author{Harumi Sakamoto\cite{address}, Tsutomu Momoi}
\affiliation{Institute of Physics, University of Tsukuba, Tsukuba
305-8571, Japan}

\author{Kenn Kubo}
\affiliation{Department of Physics, Aoyama Gakuin University,
Chitosedai, Tokyo 157-8572, Japan}
\date{\hspace*{4cm}}

\begin{abstract}
We studied ferromagnetism in the one-dimensional Hubbard model
with doubly degenerate atomic orbitals by means of the
density-matrix renormalization-group method and obtained the
ground-state phase diagrams. It was found that ferromagnetism is
stable from low to high ($0< n < 1.75$) electron density when the
interactions are sufficiently strong. Quasi-long-range order of
triplet superconductivity coexists with the ferromagnetic order
for a strong Hund coupling region, where the inter-orbital
interaction $U'-J$ is attractive. At quarter-filling ($n=1$), the
insulating ferromagnetic state appears accompanying orbital
quasi-long-range order. For low densities ($n<1$), ferromagnetism
occurs owing to the ferromagnetic exchange interaction caused by
virtual hoppings of electrons, the same as in the quarter-filled
system. This comes from separation of the charge and spin-orbital
degrees of freedom in the strong coupling limit. This
ferromagnetism is fragile against variation of band structure. For
high densities ($n>1$), the phase diagram of the ferromagnetic
phase is similar to that obtained in infinite dimensions. In this
case, the double exchange mechanism is operative to stabilize the
ferromagnetic order and this long-range order is robust against
variation of the band-dispersion. A partially polarized state
appears in the density region $1.68 \nearl n \nearl 1.75$ and
phase separation occurs for $n$ just below the half-filling
($n=2$).
\end{abstract}

\maketitle

\section{Introduction}

The Hubbard model has been employed for a  long time as a standard
model for metallic ferromagnetism of itinerant
electrons\cite{Kanamori,Gutzwiller,Hubbard}. It is, however, not
easy to show that the Hubbard model simulates a metallic
ferromagnet. Many investigations revealed that a simple
single-band Hubbard model on a  hyper-cubic lattice may not
exhibit ferromagnetism. Some additional features in the model seem
to be necessary in order to stabilize ferromagnetism. One of such
features is a special lattice or band structure. Lieb first showed
that a half-filled flat band induces a net
magnetization\cite{lieb}. Then Mielke and Tasaki introduced other
examples of flat-band ferromagnets\cite{mielke,tasaki1}. Tasaki
proved, furthermore, that a class of models with half-filled
lowest bands with finite band dispersions exhibits ferromagnetism
for sufficiently strong intra-atomic repulsions\cite{tasaki3}.
M\"uller-Hartmann found ferromagnetism in a one-dimensional system
where the band dispersion has two minima\cite{muller-hartmann}. It
was also shown by perturbational as well as numerical methods that
partially filled bands in such systems can realize
ferromagnetism\cite{penc,sakamoto,pieri,Daul}.

Another important factor realizing ferromagnetism was discussed to
be degeneracy of atomic orbitals\cite{Slater,Zener,Van}. Actually
magnetic carriers of typical metallic ferromagnets originate from
the degenerate 3$d$-orbitals, though they are mixed with $s$- or
$p$-orbitals. In the presence of degenerate orbitals, Hund
coupling favors local triplet spin configurations of two electrons
occupying  different orbitals at the same site. Following
mechanisms were proposed  which lead to bulk ferromagnetism from
intra-atomic Hund coupling.

Zener proposed the so-called {\it double exchange} mechanism
originally for doped manganites, which works when the electron
density per site is not an integer\cite{Zener,Anderson-H}, e.g.,
the density of electrons per site ($n$) is between $I$ and $I+1$.
If electrons are strongly correlated, each lattice site is
occupied by either $I$ or $I+1$ electrons, and Hund coupling
forces their spins to be parallel in an atom. The magnitude of the
formed atomic spin is $I/2$ or $(I+1)/2$ for $I\le N_{\rm d}-1$,
and $N_{\rm d}-I/2$ or $N_{\rm d}-(I+1)/2$ for $I\ge N_{\rm d}$,
where $N_{\rm d}$ denotes the degree of degeneracy. An electron
can hop from  a site occupied by $I+1$ electrons to a site
occupied by $I$ electrons. Since the hopping matrix is diagonal
with respect to the spin indices,  an electron can  hardly hop
between  sites whose spins are directed antiparallelly, while it
can freely propagate among sites with parallel spins. As a result,
ferromagnetic spin alignment that lowers the kinetic energy is
favored.

Another mechanism leading to ferromagnetism is most effective when
the electron density per site is an integer, $n=I$, where $I$ is
an integer which satisfies $1\le I\le N_{\rm d}-1$ or $N_{\rm
d}+1\le I \le 2N_{\rm d}-1$\cite{Roth}. Each site is occupied by
$I$ electrons in the strong coupling regime and hoppings of
electrons occur as virtual processes. The intermediate state of a
second order process has the lowest energy if the spins are
aligned parallelly. This leads to an effective ferromagnetic
exchange interaction that induces orbital ordering at the same
time. Van Vleck argued that the  ferromagnetic exchange due to
virtual hoppings might be the origin of the metallic
ferromagnetism in Ni\cite{Van}.

Though the mechanisms favoring ferromagnetism can be qualitatively
understood as above, it is far from trivial whether ferromagnetic
long-range order occurs in bulk systems. We need to examine
whether the models with orbital degeneracy really simulate
ferromagnets by reliable methods.  For this purpose, we study  the
simplest model with orbital degeneracy,  the  Hubbard model with
doubly degenerate atomic orbitals. The model is described by the
Hamiltonian
\begin{eqnarray}
H &=& -\sum_{i<j} t_{ij} \sum_{m=A,B} \sum_{\sig=\uparrow,\downarrow}
        (c_{im\sigma}^{\dagger}c_{jm\sigma}+{\rm H.c.}) \nonumber \\
& & +U\sum_{i}\sum_{m=A,B} n_{im\uparrow}n_{im\downarrow}
   +U'\sum_{i}\sum_{\sigma=\uparrow,\downarrow}n_{iA\sigma}n_{iB-\sigma}
\nonumber \\
& & +(U'-J)\sum_{i}\sum_{\sigma=\uparrow,\downarrow}
         n_{iA\sigma}n_{iB\sigma}  \nonumber \\
& & -J\sum_{i}
        (c_{iA\uparrow}^{\dagger}c_{iA\downarrow}
         c_{iB\downarrow}^{\dagger}c_{iB\uparrow}+{\rm H.c.})
         \nonumber \\
& & -J'\sum_{i}
        (c_{iA\uparrow}^{\dagger}c_{iA\downarrow}^{\dagger}
         c_{iB\uparrow}c_{iB\downarrow}+{\rm H.c.}),
\label{eq:deghub}
\end{eqnarray}
where $c_{im\sigma}^{\dagger}$ $(c_{im\sigma})$ creates
(annihilates) an electron at the site $i$ with spin
$\sigma\,(=\uparrow,\,\downarrow$) in the orbital $m\,(=A,\,B)$.
Here we assumed that only the hoppings between the same orbitals
exist, and the Coulomb interaction works only between electrons on
a same site. The interaction parameters  $U$, $U'$, $J$ and $J'$
represent the intra- and the inter-orbital repulsion, the
exchange (Hund-rule) coupling, and the  pair hopping,
respectively. Particle-hole symmetry allows us to study only the
case with $0\le n\le 2$.

Quite a few previous studies exist on the doubly degenerate
Hubbard model. One must be careful in comparing them, since
different assumptions were employed for the interaction
parameters. Many assumed that $J'=0$. In fact, these parameters
are dependent on each other if we calculate them from orbital wave
functions. The relation
\begin{equation}
J = J'
\label{eq:rel1}
\end{equation}
holds when  the orbital wave functions are real.
In addition,
the interaction parameters $U,$ $U'$ and $J$ satisfy the relation
\begin{equation}
U = U'+2J
\label{eq:rel2}
\end{equation}
for $e_{g}$ orbitals which is relevant for the cubic symmetry.
This Hamiltonian has the spin SU(2) and charge U(1) symmetries.
In addition, orbital degrees of freedom have U(1) rotational
symmetry since
\begin{equation}
[H,\sum_i \tau_i^y]=0.
\end{equation}
Here the orbital pseudo-spin operator is defined as
$\tau_i^y=\frac{1}{2}\sum_{\sigma=\uparrow,\downarrow} {\bf
C}_{i\sigma}^\dagger \sigma_i^y {\bf C}_{i\sigma}$, where
$\sigma_i^y$ denotes Pauli's matrix and ${\bf
C}_{i\sigma}^\dagger=(c_{iA\sigma}^\dagger,
c_{iB\sigma}^\dagger)$. The last symmetry holds if the coupling
parameters satisfy the relations (\ref{eq:rel1}) and
(\ref{eq:rel2}), and it could be easily seen in the effective
Hamiltonian derived in the strong coupling limit\cite{MomoiK}.
Because of this rotational symmetry, a rigorous argument for
one-dimensional systems excludes oscillating orbital LRO
corresponding to $\tau_i^z$-correlations\cite{Momoi}. In the limit
of $J(=J')=0$, the symmetry of orbital degrees of freedom becomes
SU(2) invariant and the whole Hamiltonian has the high symmetry
SU(4).

We assume the relations (\ref{eq:rel1}) and (\ref{eq:rel2}), and
assume all interaction parameters to be positive throughout this
work. In realistic situations the parameters satisfy $U > U' > J
$, but we study the strong Hund coupling case $J \ge U'$ as well,
for theoretical interest. Let us explain  the level structure of a
doubly occupied  atomic site since it rules  the physics  of the
model (\ref{eq:deghub}). The spin-triplet states, where two
electrons occupy different orbitals, have the lowest energy
$U'-J$. In one of  three spin-singlet states, electrons occupy
different orbitals and the energy is $U'+J$. In the other two
singlets with the energies $U-J'(=U'+J)$ and $U+J'(= U'+3J)$,
electrons occupy the same orbital. In the strongly correlated
regime where  $U' > J \gg |t_{ij}|$,  almost all sites are empty
or singly occupied  for $n<1$, while they are singly or doubly
occupied for $n>1$ and spins in a doubly occupied site form a
triplet. Thus the ground state properties for $U'>J$ strongly
depend on  whether $n$ is less than, equal to or more than unity.
If $J$ is larger than $U'$, a doubly occupied site has negative
interaction energy and hence electrons tend to form bound triplet
pairs. In this case, sites are empty or doubly occupied for
$0<n<2$ if $N_{\rm e}$ is even. If $N_{\rm e}$ is odd, there can
be an unpaired electron, which will cause  a strong finite-size
effect.

Roth examined the model in three dimensions in the quarter-filled
case ($n=1$)\cite{Roth}. She found that the ferromagnetic ground
state has an orbital superlattice structure in which two
sublattices are occupied by electrons of different orbitals when
the interaction is strong. This kind of ground state was studied
by using an effective Hamiltonian in the strong coupling
limit\cite{Kugel,Cyrot,Inagaki} and   by means of the mean field
theory for general electron density\cite{Inagaki-K}. Several exact
diagonalization studies were performed for small clusters in  one
dimension\cite{kusakabe3,kusakabe2,Gill,Kuei,Hirsch}. Most of them
investigated  the quarter-filled case ($n=1$) and found the
ferromagnetic ground state  for $U'\nearg J$. Hirsch studied the
less-than-quarter-filled case and found the ferromagnetic ground
state. Their results, however, depend strongly on the boundary
conditions and the number of lattice sites. There are rigorous
proofs for the ferromagnetic ground state in strong coupling
limits in one dimension\cite{Kubo,kusakabe2,Shen}. One must be
careful for that these proofs are valid  in different limits of
strong coupling. For the strong Hund coupling case
 ($U\rightarrow \infty$ and $U'>J\rightarrow \infty$),
existence of ferromagnetism is proved for
$1<n<2$\cite{Kubo,kusakabe2}. In the special limit $J=U'
\rightarrow \infty$ and $U\rightarrow \infty$, ferromagnetism is
stable  for $0<n<2$\cite{kusakabe2}. Shen obtained a rather
general result that the ground state is fully spin-polarized for
any $n$ between 0 and 2 except for 1 if $U=\infty$, and $U'(>0)$
and $J=J'(>0)$ are finite\cite{Shen}. The present model
(\ref{eq:deghub}) was also studied in infinite dimensions using
the dynamical mean-field theory\cite{MomoiK,Held}. Under the
assumptions (\ref{eq:rel1}) and (\ref{eq:rel2}), the ferromagnetic
ground state with and without orbital order were found for $n=1$
and $n>1$, respectively, while ferromagnetism  was not found for
$n<1$ up to the interactions of $U=40$ and $J\le
U'=20$\cite{MomoiK,difference}. B\"unemann {\it et al.} studied
ferromagnetism in the three-dimensional two-band model with rather
realistic DOS for Ni using a Gutzwiller approximation. Their
results showed that increase of DOS at the Fermi level could
stabilize ferromagnetism for electron-doped cases
($2>n>1$)\cite{Bunemann}.

In this paper, we studied the ground state of the model
(\ref{eq:deghub}) in one dimension under open boundary condition,
using the density-matrix renormalization group (DMRG)
method\cite{white}. This method enables us to study larger systems
than those studied previously. Owing to the open boundary
condition, we found remarkably little size-dependence. If we apply
the periodic boundary condition, electrons can exchange their
positions turning around the chain and it causes very large
size-dependence of the ground state (e.g. even-odd oscillations).
Preliminary results of this study were reported previously in
Ref.\ \onlinecite{Kubo2}.

This paper is constructed as follows: In Section \ref{sec:method},
we briefly review the DMRG method. In Section
\ref{sec:main-result}, we study the present model with only
nearest-neighbor hoppings and report our numerical results, such
as magnetic phase diagrams. Ferromagnetic phases appear for a wide
parameter region and for all electron density ($0<n<2$). In most
of density region, the ferromagnetic states are fully polarized,
whereas partially polarized ferromagnetic states are found in a
small region ($1.68\le n \le1.75$). We also found coexistence of
metallic ferromagnetism and quasi-long-range order of triplet
superconductivity when the Hund coupling is stronger than the
inter-orbital Coulomb repulsion ($J>U'$). Near the half-filling
case ($1.75<n<2$), phase separation occurs, and the paramagnetic
(Haldane) phase and ferromagnetic one coexist. In Section
\ref{sec:sub-result}, we examine the stability of the
ferromagnetic states obtained in Section \ref{sec:main-result},
adding perturbation to the density of states. The metallic
ferromagnetic order for more than quarter-filling ($n > 1$) is
stable against this perturbation, but, for less than
quarter-filling ($n<1$), it easily becomes unstable by this
variation. Finally, we give summary and discussion in Section
\ref{sec:summary}. Appendixes contain rigorous results in the
strong coupling limit. Using charge and spin-orbital separation,
we show that the ferromagnetic spin state for less than
quarter-filling is the same as one at quarter-filling. The
ferromagnetism at less than quarter-filling is created by the
exchange interaction produced by virtual second-order hopping
processes.

\section{Method}
\label{sec:method}

We obtained the ground state of the model (\ref{eq:deghub}) with
up to 62 sites using the density matrix renormalization group
(DMRG) method\cite{white}. The method is standard and we will not
describe it in detail. Readers should refer to White's paper for
the  DMRG method. We note only several points particular to this
study. We employed the open boundary condition throughout this
paper. In the DMRG calculations, we always kept rotational
symmetry in the spin space so that the obtained ground state is an
eigenstates of the total spin\cite{sakamoto}.

We used the finite-system method of DMRG  to obtain the
ground-state phase diagrams of the model with up to 16 sites. This
method is superior to the infinite-system method in terms of
accuracy of the ground-state quantities of finite systems. Another
merit of this method is that the final result does not depend so
strongly on the ground state of the initial small system and on
the path of the renormalization process as in the infinite-system
method. Repeated sweeps of the renormalization processes in the
finite-system method can usually remedy the failure due to a wrong
initial state. Near the phase boundary between a paramagnetic
phase and a ferromagnetic one with saturated magnetization,
however, the final results sometimes depend on the choice of the
renormalization paths. For example, two different choices of the
paths $(N,N_{\rm e})=(4, 2) (6, 3) (8, 4) (10, 5) (12, 6) (14, 7)
(16, 8)$ and $(N,N_{\rm e})=(4, 2) (6, 4) (8, 4) (10, 6) (12, 6)
(14, 8) (16, 8)$ give a perfect ferromagnetic ground state and a
paramagnetic one, respectively, at $(N, N_{\rm e}) = (16, 8)$ for
$U'=J=2|t|$, where $N_{\rm e}$ and $N$ are the number of electrons
and lattice sites, respectively. Since the transition is of first
order, the energy levels of a paramagnetic state and a
ferromagnetic one naturally exist very close to each other near
the phase boundary. To study the phase diagram carefully, we
searched the ground state through many renormalization paths.
Finally, we regard the lowest energy state as the ground state.
Hence, we need to try several paths and initial states and compare
their results in order to obtain a reliable phase diagram as long
as the number $m_{\rm D}$ of reserved eigenstates of the density
matrix is finite. The maximum number of $m_{\rm D}$ was chosen to
be 200 in this study for technical reasons.

In order to calculate longer correlations we also studied the
systems with up to 62 sites using the infinite-system DMRG method.
Results of short chains were obtained by using the exact
diagonalization method.

\section{Results for the nearest-neighbor model}
\label{sec:main-result}

In this section, we present the results for the model with only
nearest neighbor hoppings $(t_{ij}=t\delta_{i,j\pm 1})$.

\subsection{Quarter-filled system}

\begin{figure}[tbh]
\includegraphics[width=7.5cm]{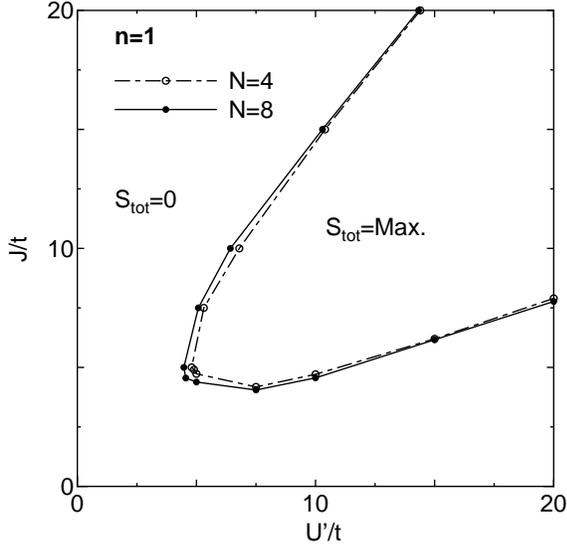} \caption{Ground-state
phase diagram at quarter-filling ($n=1$) of the model with
nearest-neighbor hoppings. Interaction parameters are set as
$U=U'+2J$ and $J=J'$. The symbol ``$S_{\rm tot}={\rm Max.}$''
denotes the perfect ferromagnetic phase, and ``$S_{\rm tot}=0$''
the paramagnetic one.} \label{fig:mgd-q}
\end{figure}

We first exhibit the results  in the quarter-filled case ($n=1$).
Figure \ref{fig:mgd-q} shows the $(U'/t,J/t)$ phase diagram of the
ground state of the quarter-filled system obtained for $N=4$ and
8. The difference between the results for $N=4$ and  $N=8$ is
amazingly small. We hence believe that the present results are
representing the phase diagram  of an infinite system, and  did
not study larger systems at this electron density. A ferromagnetic
ground state with full spin polarization appears  around $J \simeq
U'$ for $U'/t \nearg 5$. The parameter regions $J/U'\gg 1$ and
$J/U' \ll 1$ are paramagnetic. No partially polarized state was
found at this filling. The present result is similar to those by
previous studies with open boundary conditions though different
assumptions for parameters were
employed\cite{kusakabe3,kusakabe2}. In previous studies employing
periodic or anti-periodic boundary conditions, the phase
boundaries of their ferromagnetic phases depend on the system size
greatly\cite{kusakabe3,Gill,Hirsch}.

The appearance of the ferromagnetism in the region $U'>J$ can be
well understood in terms of the effective Hamiltonian in the
strong coupling limit ($U'-J \gg t$). At quarter-filling, every
site is singly occupied in this limit, and only the spin  and
orbital degrees of freedom remain. The second order effects of
virtual hoppings between singly occupied sites lead to an
effective Hamiltonian ${\cal H}_{\rm eff}$ for the spin and
orbital degrees of freedom\cite{Kugel,Cyrot,Inagaki,MomoiK}:
\begin{widetext}
\begin{eqnarray}
\lefteqn{{\cal H}_{\rm eff}  = \sum_{i=1}^{N-1} h(i,i+1)},
\label{eqn:quarter}\\
& &h(i,j) \equiv -2t^{2}\left\{\frac{4U}{U^{2}-J'^{2}}
\left(\frac{1}{4}+\tau_{i}^{z}\tau_{j}^{z}\right)
\left(\frac{1}{4}-\mbox{\bold$S$}_{i}\cdot\mbox{\bold$S$}_{j}\right) \right.
\nonumber \\
     & &  -\frac{2J'}{U^{2}-J'^{2}}
\left(\tau_{i}^{-}\tau_{j}^{-}+\tau_{i}^{+}\tau_{j}^{+}\right)
\left(\frac{1}{4}-\mbox{\bold$S$}_{i}\cdot\mbox{\bold$S$}_{j}\right)
\label{eqn:h-func} \\
             & & +\frac{2U'}{U'^{2}-J^{2}}
\left[\frac{1}{4}-\tau_{i}^{z}\tau_{j}^{z}
-2\left(\mbox{\bold$\tau$}_{i}\cdot\mbox{\bold$\tau$}_{j}
-\tau_{i}^{z}\tau_{j}^{z}\right)
\left(\frac{1}{4}+\mbox{\bold$S$}_{i}\cdot\mbox{\bold$S$}_{j}\right)\right]
\nonumber \\
             & & +\left.\frac{2J}{U'^{2}-J^{2}}
\left[\tau_{i}^{z}\tau_{j}^{z}
-\mbox{\bold$\tau$}_{i}\cdot\mbox{\bold$\tau$}_{j}
+2\left(\frac{1}{4}-\tau_{i}^{z}\tau_{j}^{z}\right)
\left(\frac{1}{4}+\mbox{\bold$S$}_{i}\cdot\mbox{\bold$S$}_{j}\right)\right]
\right\}. \nonumber
\end{eqnarray}
\end{widetext} Here $\mbox{\bold $S$}_i$ is a spin operator and
$\mbox{\bold$\tau$}_{i}=(\tau_{i}^{x},\tau_{i}^{y},\tau_{i}^{z})$
is a pseudo-spin operator representing the orbital degrees of
freedom, e.g., $\tau_{i}^{+}$ is defined as
$\sum_{\sig}c_{iA\sig}^{+}c_{iB\sig}$. According to
(\ref{eqn:h-func}) the exchange energy between the sites $i$ and
$i+1$ is $-2t^{2}/(U'-J)$, $-2t^{2}/(U'+J)$ and
$-2Ut^{2}/(U^{2}-J^{2})$ for the states $|m,\sig\ra_{i}|\barm,
\sig\ra_{i+1}$, $|m,\sig\ra_{i}|\barm,-\sig\ra_{i+1}$ and
$|m,\sig\ra_{i}|m,-\sig\ra_{i+1}$, respectively, where
$|m,\sig\ra_{i}$ denotes the state of the site $i$ occupied by an
electron of orbital $m$ with spin $\sigma$, and $\barm$ labels the
complement of $m$, e.g., ${\bar A} = B$. As a result, ${\cal
H}_{\rm eff}$ favors ferromagnetism with an orbital
antiferromagnetic superlattice structure, where electrons on two
sublattices occupy different orbitals. Numerical results of the
four sites system described by ${\cal H}_{\rm eff}$ exhibited that
the ground state of ${\cal H}_{\rm eff}$ is ferromagnetic for
$J>0.35U'$. This agrees with our observation that the lower
boundary of the ferromagnetic phase in Fig.\ \ref{fig:mgd-q}
approaches the line $J=0.35U'$ for large $U'/t$. We may consider
that the system is well described by ${\cal H}_{\rm eff}$  for
$U'/t \nearg 10$ and $U' > J$. A numerical study by two of present
authors in infinite dimensions showed that the ferromagnetism with
antiferromagnetic pseudo-spin order is stable for $J>0$ in the
strong coupling limit\cite{MomoiK}, which agrees with the result
in the classical limit of ${\cal H}_{\rm eff}$. The ferromagnetic
phase is  smaller in one dimension than that in infinite
dimensions due to stronger quantum fluctuations. In one dimension,
the antiferromagnetic orbital correlations, which supports  the
ferromagnetism, do not grow to a real long-range order but remains
as a quasi-long-range order (QLRO), since orbital degrees of
freedom have U(1) rotational symmetry and a rigorous argument
excludes oscillating orbital LRO in one dimension\cite{Momoi}.

\begin{figure}[bth]
\includegraphics[width=7.5cm]{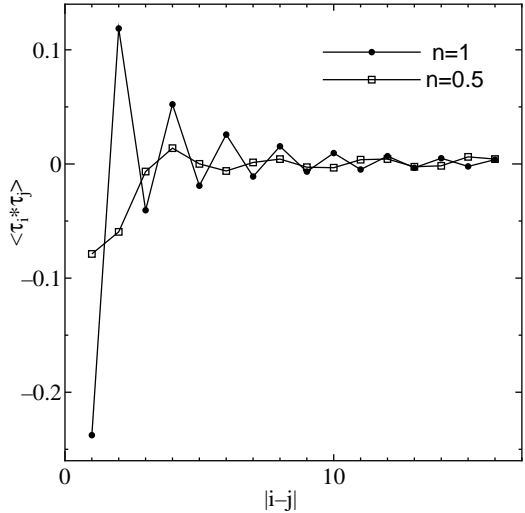} \caption{Correlation
functions of orbital ordering $\la \tau_{i} \tau_{j} \ra$ in the
ferromagnetic ground state for $J/t=15$ and $U'/t=20$ at filling
$n=1$ and $n=0.5$. The system size is 34 and $i$ is fixed at
$N/2=17$.} \label{fig:orbital}
\end{figure}

In the ferromagnetic phase at $n=1$, we found only fully polarized
states. Note that the charge and orbital degrees of freedom of
fully polarized states can be described by a single-band Hubbard
model with intra-atomic repulsion $U'-J$, where the conventional
spin degrees of freedom are replaced with the orbital
ones\cite{Shen}. Hence, we can conclude that this ferromagnetic
state has orbital antiferromagnetic QLRO for $U'-J>0$. In
addition, for general filling $n$, one can write down the
asymptotic form of the orbital correlation function in the fully
polarized ferromagnetic phase in the form
\begin{equation}\label{eq:orbital_correlation}
\langle \tau_0^z \tau_l^z \rangle \sim \cos(n\pi l)|l|^{-\alpha_S}
\end{equation}
for $U'-J>0$. The exponent $\alpha_S(=1+K_\rho)$ is obtained from
the single-band Hubbard model with coupling $U'-J$ at filling
$n$\cite{Schulz,Kawakami}. We also checked the orbital ordering by
the DMRG method and presented the results in Fig.\
\ref{fig:orbital}.

\begin{figure}[t]
\includegraphics[width=7.5cm]{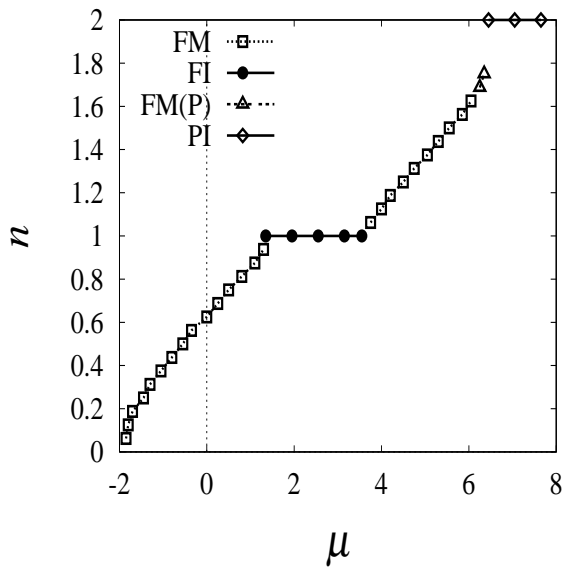}
\caption{Electron density $n$ per site as a function of the
chemical potential $\mu$ for $U'/t=15$ and $J/t=10$ for $N=16$.
The flat parts indicate insulating states. Abbreviations indicate
the characters of the ground states as FM: ferromagnetic metal,
FI: ferromagnetic insulator, FM(P): partially ferromagnetic metal
and PI: paramagnetic insulator.} \label{fig:m-in}
\end{figure}

\begin{figure}[bth]
\includegraphics[width=7.5cm]{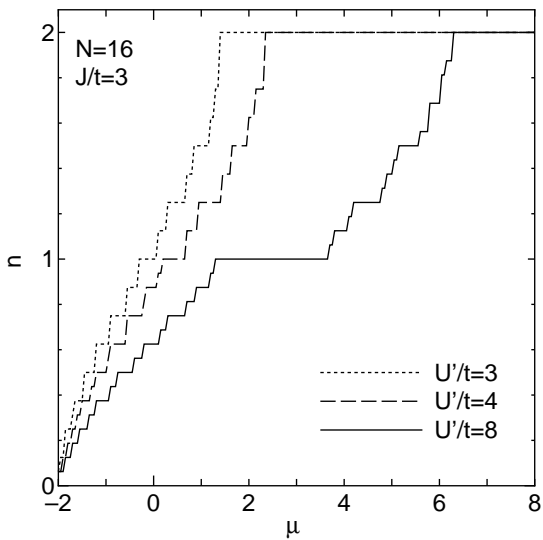}
\caption{Electron density ($n$) per site as a function of the
chemical potential ($\mu$) for $U'/t=3$, 4, and 8, and $J/t=3$ for
$N=16$. The flat parts indicate insulating states. All states are
paramagnetic.} \label{fig:n-m2}
\end{figure}

Also, from the above mapping, we can see that a metal-insulator
transition occurs in the ferromagnetic phase on the line $U'=J$,
i.e., the ferromagnetic state is insulating for $U'>J$ and
metallic for $U'\le J$. (Figure \ref{fig:m-in} shows chemical
potential dependence of the electron density for $U'/t=15$ and
$J/t=10$ with $N=16$.) On the other hand, in the paramagnetic
phase, since each band is quarter-filled and the perfect nesting
no longer exists, a metal-insulator transition may not occur with
an infinitesimal repulsion $U'-J$, which works between electrons
with the same spin in different orbitals. Indeed, for paramagnetic
states at $J=0$, Assaraf {\it et al}.\ numerically showed that a
metal-insulator transition occurs at $U_{\rm c}/t\sim
2.8$\cite{Assaraf}. We expect that, in the paramagnetic phase with
finite $J(>0)$, a metal-insulator transition may occur at a finite
positive value of $U'-J$. Figure \ref{fig:n-m2} shows numerical
results of the electron density as a function of the chemical
potential in the paramagnetic phase with $J/t=3$ and $U'/t=3$, 4,
and 8. The data for $U'/t=8$ clearly show that the quarter-filled
state is insulating. The data for $U'/t=3$ and 4 seem to indicate
that the systems are metallic at quarter filling, though it is
hard to estimate the critical point from the present calculation
because of finite-size effects.

For  $J=0$,
the effective Hamiltonian (\ref{eqn:quarter}) reduces to a model
with  $SU(4)$ symmetry.
It is known that the ground state  of this model is
a spin-singlet state $(S=\tau=0)$
in one dimension\cite{Sutherland}.
The paramagnetic state obtained for $ 0.35U' > J >0$ in Fig.\
\ref{fig:mgd-q} may be interpreted as
an extension of this $SU(4)$ symmetric state.

We find that, for the $U'=J$ case, the ferromagnetic phase extends
down to a weak coupling region. At $U'=J$, double occupation of a
site does not cost energy and the ferromagnetic ground state is
metallic. The ferromagnetism in the strong coupling region is
stabilized by a kind of the double exchange mechanism as was
discussed rigorously for $U'=J=\infty$\cite{kusakabe2}.

The ferromagnetic phase extends to the parameter region $J>U'$. In
this region, the fully polarized states can be described by the
attractive single-band Hubbard model\cite{Shen}, whose ground
state is known to have QLRO of the pairing
correlations\cite{Bogoliubov}. Hence the ferromagnetic ground
state for $J>U'$  has  QLRO of triplet superconductivity as well.
For strong attraction ($J-U' \gg t$), all electrons are coupled
into triplet pairs if  $N_{\rm e}$ is even and the system is
described by an effective Hamiltonian ${\cal H}_{\rm b}$ in terms
of hard-core bosons with spin unity:
\begin{eqnarray}
{\cal H}_{\rm b} &=& -\sum_{i=1}^{N-1} \nyo{t}
P \left [\sum_{s=0,\pm1}({b}_{is}^{\da}{b}_{i+1 s}+\mbox{H.c.})
  +{n}_{i}^{b}+{n}_{i+1}^{b}\right]P \nonumber \\
  & & +\sum_{i=1}^{N-1} P\left[
    \nyo{J}{\mbox{\bold$S$}}_{i}^{b}\cdot {\mbox{\bold$S$}}_{i+1}^{b}
      +(2\nyo{t}-\nyo{J}){n}_{i}^{b}{n}_{i+1}^{b}\right]P,
\label{eqn:Jlarge}
\end{eqnarray}
where $\nyo{t}=2t^{2}/(J-U')$ and $\nyo{J}=2t^{2}/(J+U)$. The
summation in ${\cal H}_{\rm b}$ runs over all the nearest neighbor
sites. The creation and spin operators of a boson  at a  site $i$
are defined as
\begin{eqnarray}
b_{i\,1}^{\da} &=& c_{iA\up}^{\da}c_{iB\up}^{\da}, \nonumber  \\
b_{i\,0}^{\da} &=& \frac{1}{\sqrt{2}}
(c_{iA\up}^{\da}c_{iB\down}^{\da}
+c_{iA\down}^{\da}c_{iB\up}^{\da}), \\
b_{i\,-1}^{\da} &=& c_{iA\down}^{\da}c_{iB\down}^{\da}, \nonumber
\end{eqnarray}
and
\begin{eqnarray}
(S_{i}^{b})^{z} & = & b_{i\,1}^{\da}b_{i\,1}-b_{i\,-1}^{\da}b_{i\,-1}, \\
(S_{i}^{b})^{-} & = & \sqrt{2}(b_{i\,-1}^{\da}b_{i\,0}
                              +b_{i\,0}^{\da}b_{i\,1}), \\
(S_{i}^{b})^{+} & = & \sqrt{2}(b_{i\,1}^{\da}b_{i\,0}
                              +b_{i\,0}^{\da}b_{i\,-1}),
\end{eqnarray}
respectively. The number operator  $n_{i}^{b}$ is defined as
$\sum_{s=0,\pm 1}{b}^{\da}_{is}{b}_{is}$ and  $P$ projects out
states in which sites are occupied by more than one bosons. Since
$\nyo{J}$ is positive, ${\cal H}_{\rm b}$ does not favor
ferromagnetism and the ferromagnetic phase should not extend to
this region. We examined the phase boundary in the small system
with $N=4$ and found that the slope of the phase boundary
approaches unity for large $U'/t$. The appearance of the
ferromagnetism in the region $J \nearg U'$ is thus  interpreted as
a  smearing out of the ferromagnetic ground state at  $J=U'$ to
this region. Namely, this ferromagnetism is also caused by the
double exchange mechanism at least in the strong coupling region.
The above arguments should hold not only for the quarter-filled
case but also for the hole-doped and electron-doped cases.
Actually the slope of the phase boundary for $J>U'$ approaches
unity for large $U'/t$ also in the systems with $(N,N_{\rm
e})=(6,4)$ and $(4,6)$. Shen showed coexistence of ferromagnetism
and triplet pairing correlations in the  whole region where
$U=\infty$  and $1 < J/U' < \infty$\cite{Shen}. On the other hand
the present result shows that such a  phase appears only in a
small region  close to the line $J=U'$. This discrepancy comes
from the difference in the assumed relations between parameters.
\begin{figure}[hbt]
\includegraphics[width=7cm]{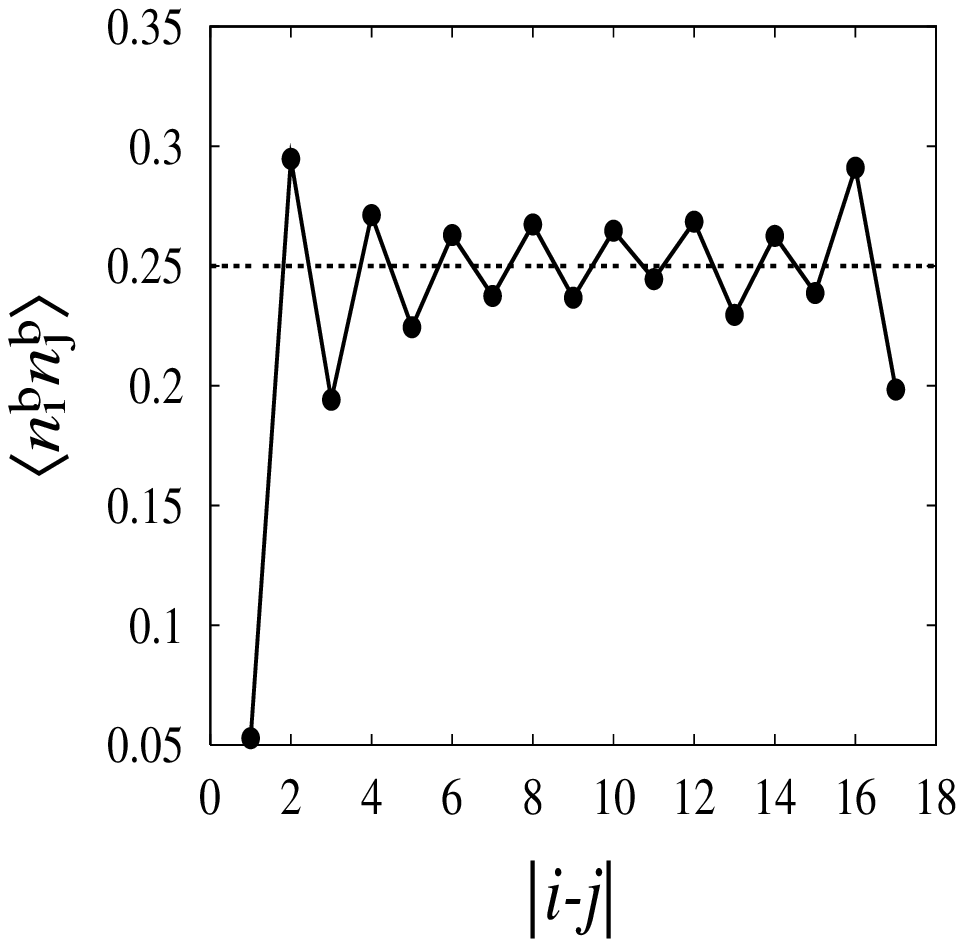}\\
\includegraphics[width=7cm]{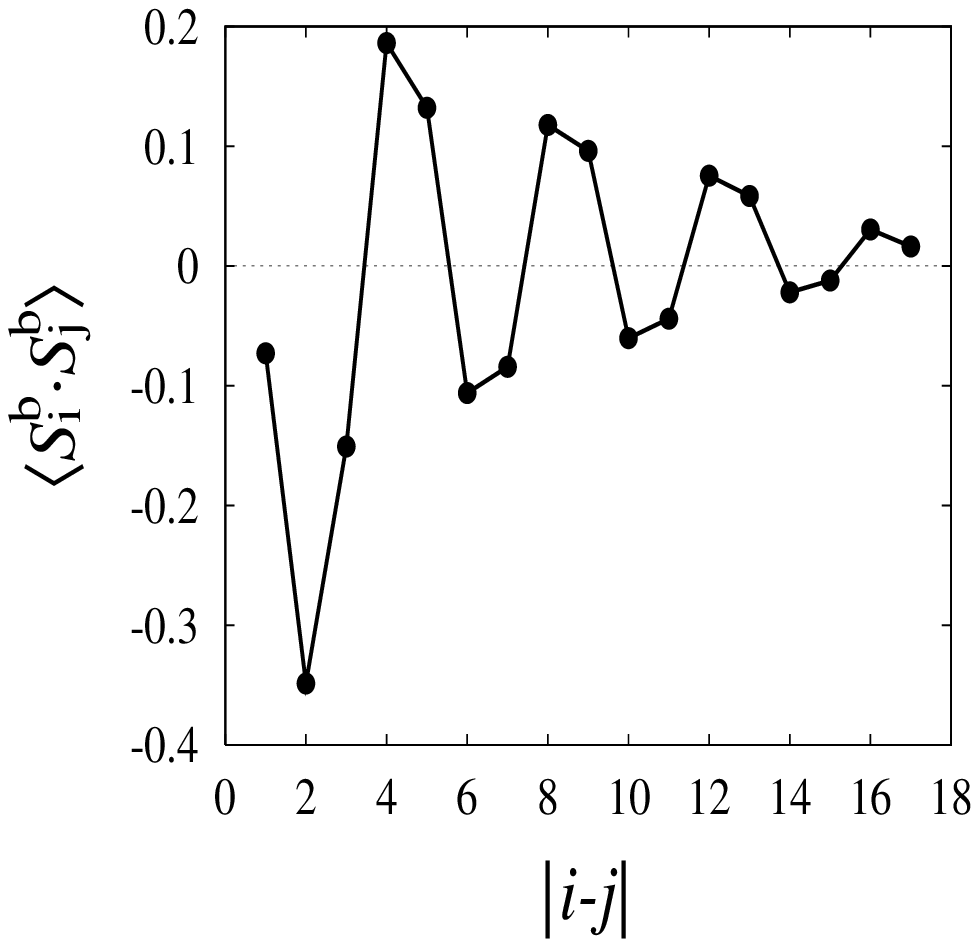} \caption{Correlation functions
$\la n_{i}^{b}n_{j}^{b}\ra$ and $\la S_{i}^{b}\cdot S_{j}^{b}\ra$
in the paramagnetic ground state  for $J/t=50$ and $U'/t=10$ at
quarter-filling. The system size is 36 and $i$ is fixed at
$N/2=18$.} \label{fig:xxz}
\end{figure}

Next, we consider the spin-singlet ground state for $J\gg U'$
where the system is described by ${\cal H}_{\rm b}$. Figure
\ref{fig:xxz} shows the density correlation function  $\la
n_{i}^{b}n_{j}^{b}\ra$ of bosons and the spin correlation function
$\la \mbox{\bold$S$}_{i}^{b}\cdot\mbox{\bold$S$}_{j}^{b}\ra$ in
the ground state of the system with 36 sites obtained by the
infinite-system DMRG method. The correlations were measured from
the center, i.e.  $i$ is fixed to be $N/2$. We observe an
even-odd oscillation  in the density correlations. This
oscillation is due to repulsive interactions between bosons.
Actually the second term in ${\cal H}_{\rm b}$ works as repulsions
since $||\nyo{J}{\mbox{\bold$S$}}_{i}^{b}\cdot
{\mbox{\bold$S$}}_{i+1}^{b}
      +2\nyo{t}-\nyo{J}||> 0$ holds when ${n}_{i}^{b}{n}_{i+1}^{b}=1$.
On the other hand,
$\la\mbox{\bold$S$}_{i}^{b}\cdot\mbox{\bold$S$}_{j}^{b}\ra$ shows
an oscillation with the period four with slow decay. This
oscillation is understood as a superposition  of the density
correlation with period two  and the antiferromagnetic spin
correlations caused by the exchange coupling  $\nyo{J}$.

\subsection{Hole- and electron-doped systems}

\begin{figure}
\includegraphics[width=7.5cm]{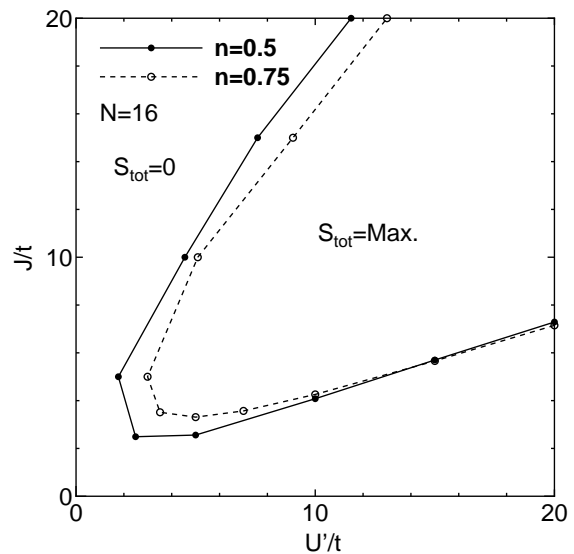}
\caption{\label{fig:mgd-h}Ground-state phase diagram for densities
$n=0.5$ and $0.75$ in the $N=16$ system. Interaction  parameters
satisfy $U=U'+2J$ and $J=J'$.}
\end{figure}

Next we study the hole-doped case ($n<1$), where holes are doped
into the  quarter-filled system. Figure \ref{fig:mgd-h} shows the
$(U'/t, J/t)$ phase diagram of the ground states for $n=0.5$ and
$0.75$ in the system with $N=16$. A ferromagnetic phase appears
for a wide parameter region and the magnetization is saturated in
the whole phase. Size-dependence among the results for $N=8$, 12
and $16$ was very weak, as in the case of quarter-filling.
Comparing with the result for $n=1$ we note that the ferromagnetic
phase has expanded in the weak-coupling region down to $U'/t\simeq
3$ as well as to a  region where $J \ge U'$.

We note that the lower boundary of the ferromagnetic phase for
large $U'/t\,(>J/t)$ almost  agrees  with that for
quarter-filling. This  result is a manifestation of the separation
of  the charge and spin-orbital degrees of freedom in one
dimension. We can derive this result by using the same argument as
that employed by Ogata and Shiba for the single-band Hubbard
model\cite{Ogata}. In the limit $U'-J=\infty$, the motion of
electrons  is  described by that of   spinless fermions on a
single chain, and the ground-state wave function is decomposed
into a spinless fermion (charge) part and a spin-orbital part. The
spin-orbital part is  described in terms of  spins and
pseudo-spins in a {\it squeezed} system with $N_{\rm e}$ sites
where empty sites are omitted. Interactions among the spins and
pseudo-spins  are caused by virtual hopping processes  and
described by the following  Hamiltonian (see Appendix
\ref{sec:hole-effec})
\begin{equation}
\label{eqn:hole-eff0}
{\cal H'}_{\rm eff}  =   n\left( 1-\frac{\sin 2n\pi}{2n\pi}\right)
{\cal H}_{\rm eff},
\end{equation}
where ${\cal H}_{\rm eff}$ is  the  Hamiltonian given  by Eq.\
(\ref{eqn:quarter}) where the site-labels of the spin-orbital
quantities are those of  the squeezed system. Since ${\cal
H'}_{\rm eff}$ differs  from ${\cal H}_{\rm eff}$ only by a
multiplicative factor, it leads to the $(U'/t, J/t)$ phase diagram
identical to that for  quarter-filling in the strong coupling
limit. This argument explains the present result that the lower
boundaries of the ferromagnetic phases for $n=0.5$, $0.75$, and
$1$ agree very well for $U'/t\nearg 10$. We may conclude  that the
origin of the ferromagnetism in this region is the effective
exchange coupling due to virtual hoppings. Orbital degrees of
freedom are also similar to those at quarter-filling except for
the difference of the characteristic wavelength. This behavior can
be seen in the orbital correlation function in the ferromagnetic
phase obtained by the DMRG method (Fig.\ \ref{fig:orbital}) and
the asymptotic form given by Eq.\ (\ref{eq:orbital_correlation}).
Note that the exponent $\alpha_S$ in Eq.\
(\ref{eq:orbital_correlation}) does not depend on the filling $n$
in the strong coupling limit.

The above result greatly depends on the fact that the wave
function is decoupled to the charge and  spin-orbital parts in the
limit $U'-J=\infty$, and that the ground-state energy is
independent of the spin-orbital degrees of freedom in the first
order of $t$. It should be noted that this is not the case in
either  higher dimensions or one-dimensional models with
far-neighbor hoppings. In these cases  the motion of electrons
leads to effective spin-orbital interactions in the first order of
$t$, as in the case of Nagaoka's ferromagnetism. In fact,
ferromagnetism was not found for $n<1$ in infinite
dimensions\cite{MomoiK,difference}. We will discuss the effects of
far-neighbor hoppings in Section \ref{sec:sub-result}.

It is remarkable that the ferromagnetism appears for  rather  weak
Hund-rule coupling, that is, $J/t\simeq 3$ for $U'/t\simeq 5$.
Hirsch argued based on a  diagonalization study of finite-size
systems that the Hund-rule coupling is not effective enough to
realize ferromagnetism in systems with low density $(n<1)$ and
that ferromagnetic exchange interaction between different sites
are necessary to explain ferromagnetism in systems such as
Ni\cite{Hirsch}. The present result, however, shows that a
moderate Hund-rule coupling is sufficient to realize the
ferromagnetism in the one-dimensional model with nearest-neighbor
hoppings.

\begin{figure}
\includegraphics[width=7.5cm]{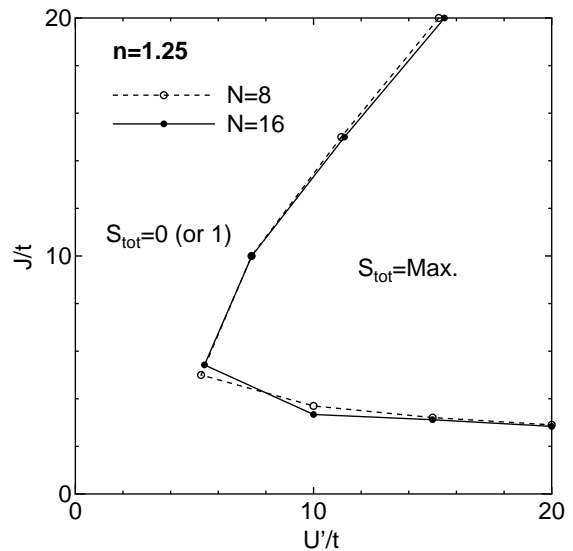}
\caption{\label{fig:mgd-e}Ground-state phase diagram at density
$n=1.25$ for the system with 8 and 16 sites. Interaction
parameters satisfy $U=U'+2J$ and $J=J'$. In the case of $N=8$,
paramagnetic states show small magnetization $S_{\rm tot}=1$
because of a finite-size effect under open boundary condition.
(See text.)}
\end{figure}

Finally we examined the electron-doped case where $n>1$. The phase
diagram for  $n=1.25$ in the systems with $N=8$ and 16 is shown in
Fig.\ \ref{fig:mgd-e}. Size-dependence was very weak, as in the
case of quarter-filling. In the systems with $N_{\rm e}=4m+2$,
where $m$ is an integer, ground states in the weak-coupling region
show small magnetization $S_{\rm tot}=1$ instead of $S_{\rm
tot}=0$. This weak spin polarization occurs when one-electron
states at the Fermi level are half-filled, i.e. $N_{\rm e}=4m+2$,
under the open-boundary condition. The Hund coupling aligns the
two electrons at the Fermi level parallel and produces the $S_{\rm
tot}=1$ ground state. We regard the appearance of $S_{\rm tot}=1$
weak polarization as a finite-size effect and not as an evidence
of unsaturated ferromagnetism in bulk systems.
In the ferromagnetic phase, the magnetization is always saturated
at this electron density. The ferromagnetism appears in a larger
region than in the quarter-filled case in particular for small
$J/t$ and large $U'/t$. This enhancement of the stable
ferromagnetic state is a result of the double exchange mechanism,
which works effectively for  $1<n<2$. The appearance of
ferromagnetism for $U>U'>J\gg t$ is expected from a  rigorous
proof which holds in the limit where $U>U'>J \rightarrow
\infty$\cite{Kubo}. Our numerical results agree with  this
argument and assure that the ferromagnetism extends to a finite
parameter region.

We show the electron density $n$ as a function of the chemical
potential $\mu$ for $U'/t=15$ and $J/t=10$ in Fig.\ \ref{fig:m-in}
for the system with 16 sites. For $n>1$ and $n<1$, the
ferromagnetic ground state is metallic, since the compressibility
${\rm d}n/{\rm d}\mu$ is finite. For $n=1$ and $2$, the value of
${\rm d}n/{\rm d}\mu$ is vanishing and hence the ground state is
insulating. The system is a fully polarized ferromagnetic
insulator for the quarter-filling case. On the other hand, the
ground state is paramagnetic at the half-filling ($n=2$) where
triplet electron pairs occupy all the lattice sites. Triplet pairs
on nearest neighbor sites are interacting antiferromagnetically
due to the second-order effect of virtual hoppings. As a result
the ground state at $n=2$ should be a Haldane singlet with an
excitation gap. There is a jump in the density between $n=1.75$
and $2$ in Fig.\ \ref{fig:m-in}, which  implies the occurrence of
phase separation. The occurrence of phase separation was found
also in  infinite dimensions\cite{Kubo2}.

\begin{figure}
\includegraphics[width=7.5cm]{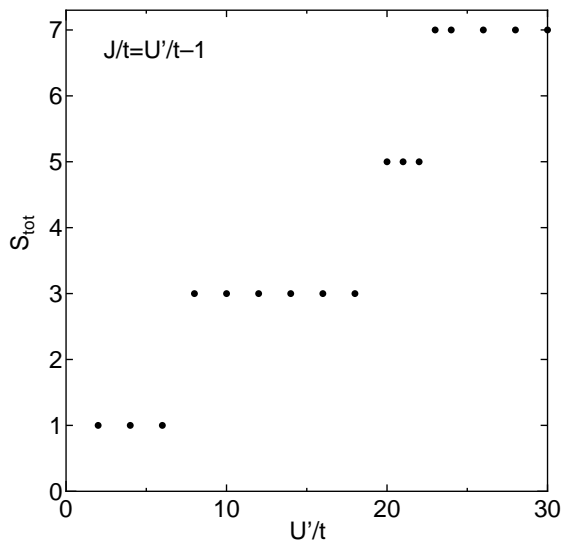}
\caption{Total spin of the ground state as a function of $U'/t$ in
the system where $J/t=U'/t-1$ and $(N,N_{\rm e})=(8,14)$.}
\label{fig:partial}
\end{figure}

We found partially polarized ground states in a small region of
density  where $1.6875 \le n \le 1.75$. As the density increases
and approaches the half-filling,  the antiferromagnetic
interactions between the triplet pairs become operative and
suppress the ferromagnetism due to the double exchange mechanism.
This may cause the appearance of the  partially polarized states.
We show the magnetization of a small system where  $(N, N_{\rm
e})=(8, 14)$ as a function of $J$ in Fig.\ \ref{fig:partial}. The
figure  shows that the partially polarized state  appears in  a
finite range of the parameter, i.e., $8 < U'/t < 23$ with
$J/t=U'/t-1$. This result may imply that the transition from a
paramagnetic to ferromagnetic ground state is of the second order
for $n\nearl 2$ in contrast to the discontinuous change to a fully
polarized state observed in other lower density region.

\section{Effect of the band-dispersion}
\label{sec:sub-result}

\begin{figure}
\includegraphics[width=7.5cm]{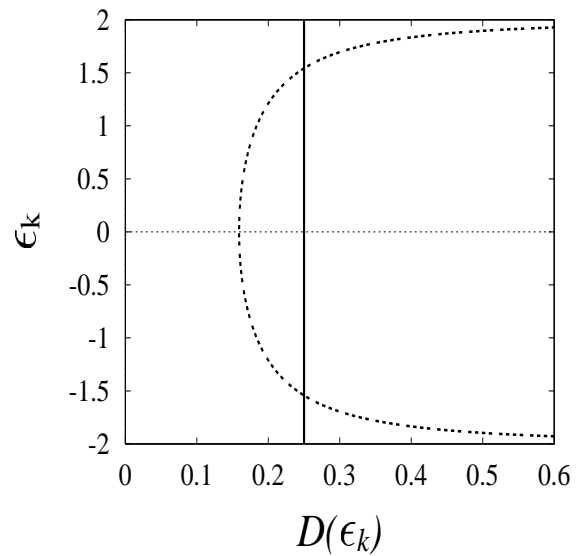}
\caption{Density of states per site and spin in the
nearest-neighbor model (dotted line) and in the linear-band model
(solid line).} \label{fig:dos}
\end{figure}

It is widely known  that the profile of the density of states
(DOS) of the band  plays an important role in the realization of
itinerant ferromagnetism. In the Hartree-Fock theory of the
Hubbard model, for example, the ferromagnetism is stabilized if
$UD_{\rm F}>1$ where $D_{\rm F}$ denotes  the DOS  at the Fermi
level. In fact, the whole profile of the DOS affects the stability
of the ferromagnetism\cite{Kanamori,Sakurai}. The DOS of the
nearest-neighbor  model  in one dimension, which we have studied
above, has a  minimum at the center of the band  and diverges at
the band-edges. These features are quite different from  those of
DOS in  higher dimensions. In this section, we examine the effect
of the band-dispersion (or DOS) on ferromagnetism, studying a
one-dimensional model with far-neighbor hoppings. The model we
employ has a linear dispersion of the band as
\begin{equation}
\epsilon_{k} = \left\{
\begin{array}{ll}
-2t+(4tk/\pi) & (0 < k < \pi) \\
-2t-(4tk/\pi) & (-\pi < k < 0)
\end{array}
\right.
\label{eqn:line}
\end{equation}
instead of $-2t\cos k$ of the nearest-neighbor model, and hence
its  DOS  is independent of the band energy
(see Fig.\ \ref{fig:dos}).
For this model, the hopping integrals are calculated
from the dispersion (\ref{eqn:line}) by
\begin{equation}
t_{ij} = \frac{2}{N+1}\sum_k \epsilon_k \sin(ki)\sin(kj),
\end{equation}
where the Fourier transform of $c_{im\sig}^{\da}$ is given by
\begin{equation}
c_{k m \sig}^{\da}  =  \sqrt{\frac{2}{N+1}}\sum_{i=1}^{N}\sin(ki)
                          c_{i m \sig}^{\da}
\label{eqn:fourier}
\end{equation}
and $k = \ell\pi/(N+1)$ with $\ell=1,2,..,N$. Due to the
electron-hole symmetry of the dispersion (\ref{eqn:line}) the
system is bipartite, i.e. the hopping integrals $t_{ij}$ vanish
for $|i-j|=2\ell$ with integer $\ell$, and are always negative for
$|i-j|=2\ell + 1$ for positive  $t$. The hopping integral $t_{ij}$
is given by  $-t[1-(-1)^{i-j}]/\pi(i-j)^{2}$ in the limit
$N\rightarrow\infty$. We call this  model {\it linear-band} model
in the following. We study small clusters with $N=4$, 5 and 6
under the open boundary conditions, exactly diagonalizing them,
and compare the results with those of the nearest-neighbor model.

\begin{figure}
\includegraphics[width=7.5cm]{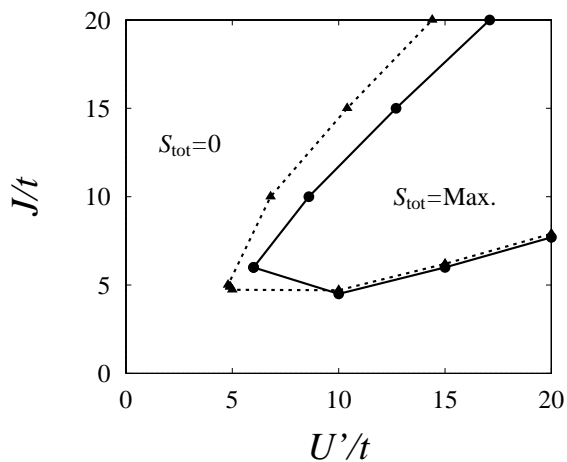}
\caption{Ground-state phase diagram at quarter-filling obtained by
the exact diagonalization of the system  with  $N=N_{\rm e}=4$.
Filled circles and filled triangles  display the phase boundary
for the linear-band  model and the nearest-neighbor model,
respectively.} \label{fig:lmgd-q}
\end{figure}

We show the ground state phase diagram for the quarter-filling
obtained from the systems with $(N,N_{\rm e})=(4,4)$ in Fig.\
\ref{fig:lmgd-q}. The phase boundaries for two models are  in good
agreement in the strong coupling region where  $U'>J$, while the
ferromagnetism is slightly suppressed in the linear-band model in
the weak-coupling region where $J\ge U'$. The ferromagnetism is
assisted by antiferromagnetic orbital QLRO when  $U'>J$. The
long-range hoppings  in the linear-band model do not destroy the
orbital correlations since $t_{ij}$ satisfies the bipartite
condition. The present result suggests  that the phase  boundary
between the ferromagnetic and paramagnetic phases does not depend
strongly on the spatial dependence of the effective exchange
interactions.

\begin{figure}
\includegraphics[width=7.5cm]{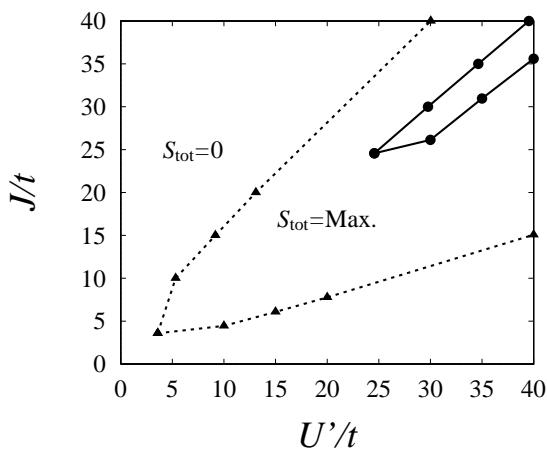}
\caption{Ground-state phase diagram for the system with
  $(N,N_{\rm e})=(6,4)$.
Filled circles and filled triangles  display the phase boundary
for the linear-band  model and the nearest-neighbor model,
respectively.} \label{fig:lmgd-h}
\end{figure}

Next we show the  phase diagram for $(N,N_{\rm e})=(6,4)$ in Fig.\
\ref{fig:lmgd-h} as an example of  the hole-doped case. It is
striking that the ferromagnetic region for the linear-band model
is strongly suppressed compared to that for the nearest-neighbor
model. The ferromagnetic phase  appears only for $U'/t \ge 25$ and
exists  in a very narrow region along  the line $J/U'=1$. This
strong reduction of the ferromagnetic phase for $U'>J$ is
interpreted as a result of the breakdown of the separation of the
charge and spin-orbital degrees of freedom. An electron can pass
the other electrons by  far-neighbor hoppings, and they  lead to
effective interactions among  the spin and orbital degrees of
freedom in the first-order  of $t_{ij}$. The effective Hamiltonian
in the limit  $U'>J\rightarrow \infty$, which contains only
hopping terms and  prohibits  double occupancy, leads to  a unique
paramagnetic ground state ($S=0$) in the linear-band model for
$(N,N_{\rm e})=(6,4)$. Consequently, the ferromagnetic phase does
not appear for $U'>J \gg t$ in this model and occurs only in the
region $J\simeq U'$ and $U'/t\nearg 25$.

\begin{figure}
\includegraphics[width=7.5cm]{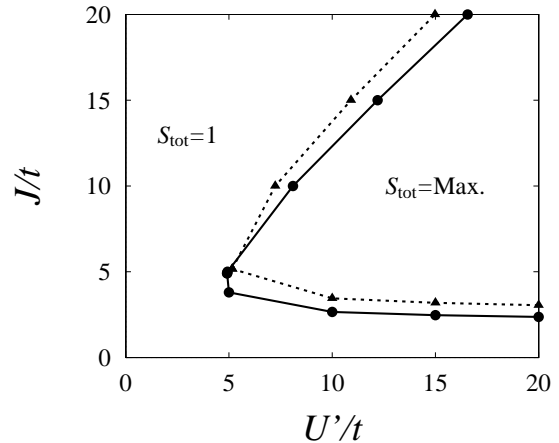}
\caption{Ground-state phase diagram for $(N,N_{\rm e})=(5,6)$.
Filled circles and filled triangles  display the phase boundary
for the linear-band  model and the nearest-neighbor model,
respectively.} \label{fig:lmgd-e}
\end{figure}

Figure \ref{fig:lmgd-e} displays the phase diagram for $(N,N_{\rm
e})=(5,6)$ as an example of the  electron-doped case. In this
case, the phase boundary for the linear-band model is similar to
that for the nearest-neighbor model. The ferromagnetism for $U' >
J \gg t$ is caused by the double exchange mechanism. Far-neighbor
hoppings are expected to favor ferromagnetism as well as
nearest-neighbor ones. We note that rigorous
proofs\cite{kusakabe2,Kubo,Shen} for the ferromagnetism in the
strong coupling limit do not hold for the linear-band model. The
present  result implies that the ferromagnetic state for $n >1$ is
robust against the variation of the  DOS and would survive  in
higher dimensions. Actually, ferromagnetic state  was found to be
stable for a wide range of interaction parameters for $n>1$ in
infinite dimensions\cite{MomoiK}.

Though we have studied only small systems, the above results
clearly show that the band-dispersion has a strong effect on the
stability of ferromagnetism for the hole-doped system. On the
other hand, it gives only a minor effect  on the ferromagnetism
for the quarter-filled and the electron-doped systems.

\section{Summary and Discussion}
\label{sec:summary}

We studied the ferromagnetism in the one-dimensional Hubbard model
with orbital degeneracy by numerical methods. We first studied
the model with only nearest-neighbor hoppings by the DMRG
calculations of clusters with up to 16 sites  and obtained the
following results:
\begin{enumerate}
\item
In the case of the quarter-filling $(n=1)$, a fully polarized
ferromagnetic phase appears in a region $U'/t \ge 4$ and $J/t \ge
4$. The ferromagnetic state is insulating for $U'>J$ and metallic
for $J\ge U'$. The ferromagnetism for $U'-J\gg t$ is caused mainly
by the second-order hopping processes. The phase boundary between
the ferromagnetic and paramagnetic phases approaches the asymptote
$J= 0.35U'$ for large $U'\,(>J)$. The ferromagnetic phase extends
considerably to a region with $J>U'$ for intermediate coupling
regime, but the phase boundary on this side  becomes parallel to
the line $J=U'$ for large $U'/t$.

\item
In the hole-doped cases ($n=0.5$ and $0.75$), metallic
ferromagnetism appears in a similar parameter region to that of
insulating ferromagnetism at the quarter-filling. The phase
boundary between the ferromagnetic and paramagnetic phases for
large $U'(>J)$ agrees with that for $n=1$, which results from
decoupling of the ground-state wave function into the charge and
spin-orbital parts in the limit $U'-J=\infty$. The ferromagnetic
phase expands to the weak-coupling region more than the
quarter-filled case.

\item
In the electron-doped case $(n=1.25)$, the ferromagnetic phase
dominates the region with $U'>J$. This ferromagnetism is metallic
and caused mainly by the double exchange mechanism. Near the
half-filled case ($1.6875 \le n \le 1.75$), this ferromagnetic
order becomes partially saturated.

\item In a large-Hund-coupling region, $J > U'$, QLRO of
triplet superconductivity coexists with the metallic ferromagnetism,
which appears for all electron densities, i.e., $0<n<1.75$.
\end{enumerate}
It is important to note that the present results for the phase
diagram showed little system-size dependence. We hence claim that
the present results represent the properties in the thermodynamic
limit.

\begin{figure}
\includegraphics[width=7.5cm]{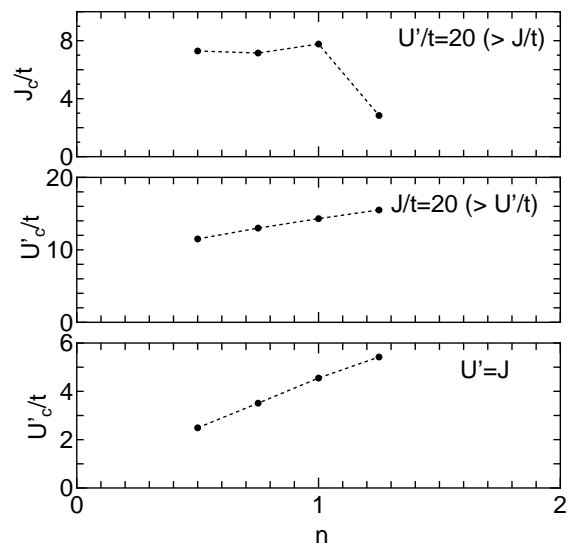}
\caption{Critical values of the interaction parameters for the
appearance of ferromagnetism as functions of the density:  $J_{\rm
c}/t$ for $U'/t=20$ (top), $U'_{\rm c}/t$ for $J/t=20$ (middle)
and $U'_{\rm c}/t$ $(=J_{\rm c}/t)$ for $U'=J$ (bottom).}
\label{fig:n-dep}
\end{figure}

The critical values of the interaction parameters for the
appearance of ferromagnetism are shown in Fig.\ \ref{fig:n-dep}.
In the case of $U'>J$, the critical Hund coupling $J_{\rm c}/t$ is
almost constant for $n\le 1$, while it suddenly drops for
$n=1.25$. Though we have not examined for other $n$ values for
$n>1$, the data suggest a clear discrepancy between $n>1$ and
$n\le 1$. For  $U'=J$ and $U'<J$, the critical inter-orbital
repulsion $U'_{\rm c}/t$ is apparently a smooth function of the
density. For the case $U'=J$,  $U'_{\rm c}/t$ seems to vanish for
decreasing $n$. It would be interesting to examine whether the
system exhibits ferromagnetism with infinitesimal interactions in
the low-density limit.

\begin{figure}
\includegraphics[width=7.5cm]{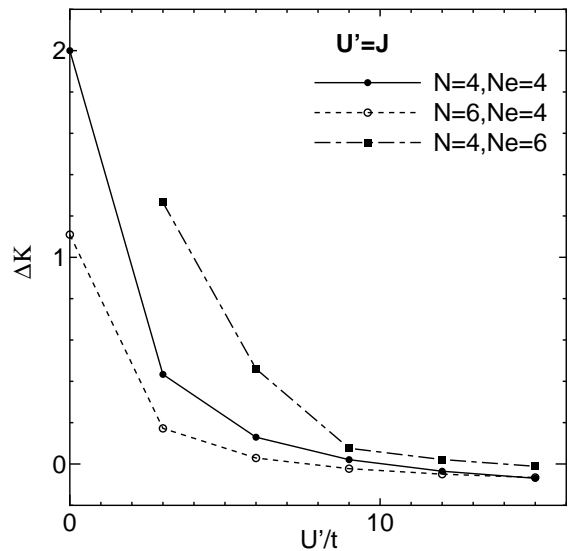}
\caption{Difference $\Delta K$ of the kinetic energy in
ferromagnetic states $K_{\rm F}$ from that in paramagnetic states
$K_{\rm P}$, i.e., $\Delta K=K_{\rm F}-K_{\rm P}$. The numbers of
system sizes and electrons are set as $(N,N_{\rm e})=(6,4)$,
$(4,4)$ and $(4,6)$.} \label{fig:deltaK-1}
\end{figure}

The double exchange mechanism causes ferromagnetism by lowering
the kinetic energy, and hence the strength of the double exchange
mechanism may be measured by the difference $\Delta K$ of the
kinetic energy in the ferromagnetic state $K_{\rm F}$ from that in
the paramagnetic one $K_{\rm P}$, i.e., $\Delta K=K_{\rm F}-K_{\rm
P}$. Largely negative $\Delta K$ implies that the double exchange
mechanism is effective. We show $\Delta K$ for $U'=J$ in the small
systems with $(N,N_{\rm e})=(6,4)$, $(4,4)$ and $(4,6)$ in Fig.\
\ref{fig:deltaK-1}. We see that $\Delta K$ is a decreasing
function of $U'/t$ in all the cases and becomes negative for large
$U'/t$. The value $U'_{0}/t $ where $\Delta K =0$ decreases with
decreasing $n$ which implies that the  double exchange mechanism
is more effective in the system with lower density for $U'=J$. The
values of  $U'_{0}/t $ themselves are, however,  much larger than
the  critical value $U'_{\rm c}/t$ for the appearance of the
ferromagnetism. For example, $U'_{0}/t \simeq 8$ and $U'_{\rm c}/t
\simeq 4$ for $(N,N_{\rm e})=(6,4)$. This result shows that the
ferromagnetism in the weak-coupling region is not caused by the
double exchange mechanism. The ferromagnetism occurs for a gain of
the interaction energy. The situation is similar in the system
with $(N,N_{\rm e})=(4,6)$ for $U'>J$ as well. The value $J_{0}/t
$ where  $\Delta K =0$ is much larger than the critical value
$J_{\rm c}/t$. For example, $J_{0}/t \simeq 7$ and $J_{\rm c}/t
\simeq 3$  for $U'/t=20$. The system-size dependence of $U'_{0}/t
$ (and $J_{0}/t$) is left for a future study; whether it agrees
with $U'_{\rm c}/t$ (and $J_{\rm c}/t$) or not in the
thermodynamic limit.

We found a partially polarized ferromagnetic state for a density
region adjoining the phase-separation region just below
half-filling, where the magnetization gradually changes depending
on the strength of interactions. In these states, ferromagnetic
spin correlation decays especially near edges of open systems. We
may need further study in order to confirm that this partial
ferromagnetism remains in the thermodynamic limit. There is also
another possibility that a partially polarized state might
apparently appear because of phase separation into a perfectly
polarized part and a paramagnetic part. Recently phase separation
is found in other systems and is thought to be a common feature of
strong correlations\cite{Yunoki,Kubo2}.

Next, we examined the effect of the band-dispersion on the
ferromagnetism using diagonalization of small clusters for a model
with far-neighbor hoppings. The ferromagnetism is greatly
destabilized at less than quarter-filling $n<1$. In this case, the
second-order effects of hoppings, which are dominating in the
nearest-neighbor hopping model, are overwhelmed by the first-order
effect. The phase diagram for the quarter-filling does not change
greatly by far-neighbor hoppings that satisfy the bipartite
condition. This result will be modified if the frustration is
introduced  by non-bipartite hopping integrals. The ferromagnetism
for the electron-doped case is quite robust against variation of
the DOS shape.

We may extract some presumptions on the ferromagnetism due to the
orbital degeneracy in two and three dimensions both from the
present results and those in infinite
dimensions\cite{MomoiK,difference}. At the quarter-filling ($n=1$)
the ferromagnetism is supported by the orbital antiferromagnetic
correlations for $U'>J$. This mechanism is common in all
dimensions and robust if the hopping integrals  satisfy the
bipartite condition. We hence expect that the insulating
ferromagnetic phase occurs in a similar parameter region in all
dimensions. The lower phase boundary will approach the line
$J=\alpha U'$ $(0.35 >\alpha >0)$ for large $U'/t$, and $\alpha $
will decrease with increasing the dimensionality. For the
electron-doped system ($n>1$), it is natural to expect that the
double-exchange mechanism works in any dimension. The metallic
ferromagnetism for $U'>J$ will appear in a similar parameter
region in any dimension for $2>n>1$, since this ferromagnetic
order is insensitive to the variation of the band-dispersion. The
situation is subtle in the case of the hole-doped system ($n<1$).
The double-exchange mechanism hardly works for $n<1$ because of
inter-orbital Coulomb repulsion. In this density region, indeed,
variation of band-dispersion strongly destabilizes ferromagnetism
in one dimension and no ferromagnetic phase was found in infinite
dimensions, though only sparse parameter points were
examined\cite{MomoiK,difference}. In three (and two) dimensions,
hence, the ferromagnetism might not appear in the simple
nearest-neighbor hopping model on  hyper-cubic lattices for $n<1$.
To confirm this argument, we need to take account of the interplay
between the electron correlations and the band-dispersion and need
to accomplish serious calculations for two- and three-dimensional
systems.

For $J>U'$, QLRO of triplet superconductivity appears, which is
supported by ferromagnetic order. In this region, a ferromagnetic
state  is stabilized  by the double exchange mechanism. However,
this ferromagnetic state is affected by far-neighbor hoppings and,
in infinite dimensions, the boundary of the ferromagnetic phase
does not extend beyond the $J=U'$ line up to the coupling
$U'=10$\cite{MomoiK}, though we (TM and KK) did not study the
$J>U'$ region. The coexistence of ferromagnetism and triplet
superconductivity for $J\ge U'$ might not appear in higher
dimensions. We expect that a two-dimensional system may be a
plausible candidate for showing ferromagnetic triplet
superconductivity.

Properties of paramagnetic ground states are of theoretical
interest as well, though we did not study them in detail. The
model has the  $SU(4)$ symmetry for $J=0$ and solved exactly in
one dimension\cite{Sutherland}. Recent numerical works clarified
the properties of the quarter-filled
system\cite{Yamashita,Frischmuth}. It will be an interesting
problem to investigate the spin and orbital correlations for
general $n$ and $J\ge 0$. From the charge and spin-orbital
separation as shown in Appendix \ref{sec:hole-effec}, we can say
that the spin degrees of freedom for $n<1$ are equal to those at
$n=1$ in the strong coupling limit. In the other paramagnetic
region for $J \gg U'$, electrons are bound to form triplet pairs.
We found strong charge correlations  for the quarter-filling, but
did not find LRO of charge density. It is an interesting problem
to study whether a charge ordering occurs and coexists with
antiferromagnetic LRO in higher dimensions.

\acknowledgments

Numerical calculations were performed on Facom VPP500 at the ISSP,
University of Tokyo, and facilities of the Information Center,
University of Tsukuba. In the numerical calculations, we partly
used subroutines of TITPACK Ver. 2 coded by H.\ Nishimori. This
work was financially supported by Grant in Aid Nod.\ 09640453,
10203202 and 11640365 from the Ministry of Education, Science,
Sports and Culture of Japan. One of the authors (K.K.) was
supported by Center for Science and Engineering Research, Research
Institute of Aoyama Gakuin University.

\appendix
\section{The Effective Hamiltonian  for $n<1$ in the Strong-coupling
Limit}
\label{sec:hole-effec}

\begin{figure}
\includegraphics[width=5cm]{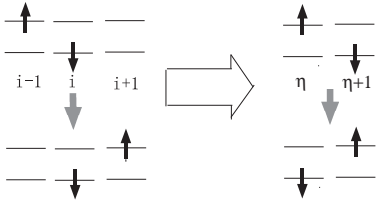}
\caption{An examples of mapping from labelling of electrons by
sites to labelling by counting only electrons from the left
($\eta$-th electron). A second-order term
$\nn_{iB\down}\nc_{i+1\,A\up}^{\da}\nc_{i-1\,A\up}$ is mapped to
the term
$S_{\eta}^{-}\tau_{\eta}^{-}S_{\eta+1}^{+}\tau_{\eta+1}^{+}$.}
\label{fig:hole-map}
\end{figure}

In this appendix, we derive the effective exchange Hamiltonian for
$n<1$ in the limit $U' -J\gg t $ in the nearest-neighbor hopping
model. We treat the hopping term in (\ref{eq:deghub}) as a
perturbation. In the unperturbed ground state, every site is
singly occupied or empty. As the first-order effect of the hopping
term, electrons hop from singly occupied sites to empty ones. The
effective Hamiltonian in the first order of $t$ is written as
\begin{equation}
H_{\rm eff}^{(1)} = -t\sum^{N}_{i=1}\sum_{m,\sig}
\left(\nc_{im\sig}^{\da}\nc_{i+1m\sig}+\mbox{H.c.}\right),
\label{eqn:hole-eff1} \\
\end{equation}
where $\nc_{jm\sig}\equiv c_{jm\sig}(1-n_{jm\,-\sig})
(1-n_{j\barm\sig})(1-n_{j\barm\,-\sig})$ is the
Gutzwiller-projected annihilation operator and $\nn_{j}$ is
defined as $\sum_{m,\sig}\nc_{jm\sig}^{\da}\nc_{jm\sig}$. In the
unperturbed ground-state subspace  $\nn_{j}$ is 0 or 1. Since the
electrons cannot exchange their positions under the open boundary
conditions, the matrix elements of $H_{\rm eff}^{(1)}$ are
independent of the spin-orbital degrees of freedom. Hence  $H_{\rm
eff}^{(1)}$  is  equivalent to a Hamiltonian  for free spinless
fermions, i.e.,
\begin{equation}
H_{\rm eff}^{(1)} = -t
\sum_{i=1}^{N}\left(a_{i+1}^{\da}a_{i}+{\rm H.c.}\right),
\end{equation}
where $a_{i}^{\da}$ is the creation operator of  a  spinless
fermion. Then the  number  operator  $\nn_{i}$ is equivalent to
$a_{i}^{\da}a_{i}$. We may write down the ground-state wave
function $|\Psi_{\rm g}\ra$ as a direct product of the ground
state of spinless fermions, $|\Phi_{\rm SF}\ra$, and a
spin-orbital wave function $|\Omega\ra$ as\cite{Ogata}
\begin{equation}
|\Psi_{\rm g}\ra = |\Phi_{\rm SF}\ra\bigotimes|\Omega\ra.
\label{eqn:decom}
\end{equation}
Here $|\Phi_{\rm SF}\ra$ is a simple Slater determinant for
spinless fermions and describes only the charge degrees of
freedom. The ground-state energy is degenerate for any
spin-orbital wave function $|\Omega\ra$ up to the first order in
terms of $t$. The second-order terms due to virtual hoppings
determine  $|\Omega\ra$. The effective Hamiltonian in the second
order is given by
\begin{equation}
H_{\rm eff}^{(2)} = H_{\rm eff}^{(2a)}
                  +H_{\rm eff}^{(2b)} \label{eqn:hole-eff2},
\end{equation}
where
\begin{equation}
H_{\rm eff}^{(2a)} = \sum_{i=1}^{N}\nn_{i}\nn_{i+1}h(i,i+1)
\label{eqn:hole-eff2a}
\end{equation}
and
\begin{eqnarray}
\lefteqn{H_{\rm eff}^{(2b)} = \sum_{i=1}^{N}\sum_{\sig}
\sum_{m}\times} \nonumber \\
& & \left[-\frac{t^{2}}{U'-J}\right.
(\nc_{i-1\,m\sig}^{\da}n_{i\,\barm\sig}\nc_{i+1\,m\sig}+\mbox{H.c.}) \nonumber \\
& & +\frac{t^{2}}{U'-J}
(\nc_{i-1\,\barm\sig}^{\da}\nc_{i\,m\sig}^{\da}\nc_{i\,\barm\sig}\nc_{i+1\,m\sig}
+\mbox{H.c.}) \nonumber \\
  & & -\frac{U't^{2}}{(U')^{2}-J^{2}}
(\nc_{i-1\,m\sig}^{\da}n_{i\,\barm\,-\sig}\nc_{i+1\,m\sig}+\mbox{H.c.})
\nonumber \\
& & +\frac{U't^{2}}{(U')^{2}-J^{2}}
(\nc_{i-1\,\barm\sig}^{\da}\nc_{i\,m\,-\sig}^{\da}\nc_{i\,\barm\sig}
\nc_{i+1\,m\,-\sig}
+\mbox{H.c.}) \nonumber \\
  & &  -\frac{Jt^{2}}{(U')^{2}-J^{2}}
(\nc_{i-1\,m\sig}^{\da} \nc_{i\,\barm-\sig}^{\da}
\nc_{i\,\barm\sig}
  \nc_{i+1\,m\,-\sig}
+\mbox{H.c.}) \nonumber \\
& & +\frac{Jt^{2}}{(U')^{2}-J^{2}}
(\nc_{i-1\,\barm\sig}^{\da}\nc_{i\,m\,-\sig}^{\da}
\nc_{i\,\barm\,-\sig}\nc_{i+1\,m\,\sig}
+\mbox{H.c.}) \nonumber \\
  & & -\frac{Ut^{2}}{U^{2}-(J')^{2}}
(\nc_{i-1\,m\sig}^{\da}n_{i\,m\,-\sig}\nc_{i+1\,m\sig}
+\mbox{H.c.}) \nonumber \\
& & +\frac{Ut^{2}}{U^{2}-(J')^{2}}
(\nc_{i-1\,m\sig}^{\da}\nc_{i\,m\,-\sig}^{\da}\nc_{i\,m\sig}\nc_{i+1\,m\,-\sig}
+\mbox{H.c.}) \nonumber \\
  & & -\frac{J't^{2}}{U^{2}-(J')^{2}}
(\nc_{i-1\,m\sig}^{\da}\nc_{i\,m\,-\sig}^{\da}
\nc_{i\,\barm\,\sig}\nc_{i+1\,\barm\,-\sig}
+\mbox{H.c.}) \nonumber \\
& & +\left.\frac{J't^{2}}{U^{2}-(J')^{2}}
(\nc_{i-1\,m\sig}^{\da}\nc_{i\,m\,-\sig}^{\da}
\nc_{i\,\barm\,-\sig}\nc_{i+1\,\barm\,\sig} +\mbox{H.c.}) \right].
\label{eqn:hole-eff2b}
\end{eqnarray}
The first term $H_{\rm eff}^{(2a)}$ represents the exchange
interactions between electrons occupying nearest-neighbor sites,
where $h(i,i+1)$ is  defined in the effective Hamiltonian
(\ref{eqn:h-func}) for the quarter-filling. The second term
$H_{\rm eff}^{(2b)}$ represents correlated hoppings between
next-nearest-neighbor sites. The effective Hamiltonian  $H_{\rm
eff}^{(2)}$ contains both the  charge and spin-orbital degrees of
freedom. We may average charge degrees of freedom in $H_{\rm
eff}^{(2)}$ out since  $|\Phi_{\rm SF}\ra$ is fixed to minimize
$H_{\rm eff}^{(1)}$. In order to accomplish the averaging we
introduce the {\it squeezed} system where  the empty sites are
omitted from the original system. Hence it is a chain with $N_{\rm
e}$ sites with spin and orbital degrees of freedom on each site.
We label the sites in the squeezed system  by $\eta$. Employing
the notation $j_{\eta}$ for the position in the original system of
the $\eta$-th electron in the squeezed system, we rewrite $H_{\rm
eff}^{(2)}$ as
\begin{equation}
H_{\rm eff}^{(2a)} =\sum_{\eta=1}^{N_{\rm e}}\sum_{i=1}^{N}
\nn_{i}\nn_{i+1} \delta_{ij_{\eta}}h(\eta,\eta+1)
\end{equation}
and
\begin{widetext}
\begin{equation}
H_{\rm eff}^{(2b)} =
\sum_{\eta=1}^{N_{\rm e}}\sum_{i=1}^{N}\sum_{\sig_{0}=\pm}\sum_{m_{0}=\pm}
\left(\nn_{i}a_{i-1}^{\da}a_{i+1}+{\rm H.c.}\right)
\delta_{ij_{\eta}}\Theta(\eta,\sig_{0},m_{0}), \label{eqn:effec-2b0}
\end{equation}
where
\begin{eqnarray}
\Theta(\eta,\sig_{0},m_{0}) &\equiv &
  \left\{\mbox{\rule{0pt}{0.6cm}}\right. -\frac{t^{2}}{U'-J}
\lef(\frac{1}{2}+\sig_{0}S^{z}_{\eta}\ri)\tau^{m_{0}}_{\eta}
\lef(\frac{1}{2}+\sig_{0}S^{z}_{\eta+1}\ri)\tau^{-m_{0}}_{\eta+1}
  \nonumber \\
& & +\frac{t^{2}}{U'-J}
\lef(\frac{1}{2}+\sig_{0}S^{z}_{\eta}\ri)
\lef(\frac{1}{2}-m_{0}\tau^{z}_{\eta}\ri)
\lef(\frac{1}{2}+\sig_{0}S^{z}_{\eta+1}\ri)
\lef(\frac{1}{2}+m_{0}\tau^{z}_{\eta+1}\ri) \nonumber \\
  & &
-\frac{U't^{2}}{(U')^{2}-J^{2}}
S_{\eta}^{\sig_{0}}\tau_{\eta}^{m_{0}}
S_{\eta+1}^{-\sig_{0}}\tau_{\eta+1}^{-m_{0}} \nonumber \\
& &
+\frac{U't^{2}}{(U')^{2}-J^{2}}
\lef(\frac{1}{2}+\sig_{0}S^{z}_{\eta}\ri)
\lef(\frac{1}{2}-m_{0}\tau^{z}_{\eta}\ri)
\lef(\frac{1}{2}-\sig_{0}S^{z}_{\eta+1}\ri)
\lef(\frac{1}{2}+m_{0}\tau^{z}_{\eta+1}\ri) \nonumber \\
  & &
-\frac{Jt^{2}}{(U')^{2}-J^{2}}
\lef(\frac{1}{2}+\sig_{0}S^{z}_{\eta}\ri)\tau^{m_{0}}_{\eta}
\lef(\frac{1}{2}-\sig_{0}S^{z}_{\eta+1}\ri)\tau^{-m_{0}}_{\eta+1}
\nonumber \\
& &
+\frac{Jt^{2}}{(U')^{2}-J^{2}}
S_{\eta}^{\sig_{0}}\lef(\frac{1}{2}-m_{0}\tau_{\eta}^{z}\ri)
S_{\eta+1}^{-\sig_{0}}\lef(\frac{1}{2}+m_{0}\tau_{\eta+1}^{z}\ri)
\nonumber \\
  & &
-\frac{Ut^{2}}{U^{2}-(J')^{2}}
S_{\eta}^{\sig_{0}}\lef(\frac{1}{2}+m_{0}\tau_{\eta}^{z}\ri)
S_{\eta+1}^{-\sig_{0}}\lef(\frac{1}{2}+m_{0}\tau_{\eta+1}^{z}\ri)
\nonumber \\
& &
+\frac{Ut^{2}}{U^{2}-(J')^{2}}
\lef(\frac{1}{2}+\sig_{0}S^{z}_{\eta}\ri)
\lef(\frac{1}{2}+m_{0}\tau^{z}_{\eta}\ri)
\lef(\frac{1}{2}-\sig_{0}S^{z}_{\eta+1}\ri)
\lef(\frac{1}{2}+m_{0}\tau^{z}_{\eta+1}\ri) \nonumber \\
  & &
-\frac{J't^{2}}{U^{2}-(J')^{2}}
\lef(\frac{1}{2}+\sig_{0}S^{z}_{\eta}\ri)\tau^{m_{0}}_{\eta}
\lef(\frac{1}{2}-\sig_{0}S^{z}_{\eta+1}\ri)\tau^{m_{0}}_{\eta+1}
\nonumber \\
& &
+\frac{J't^{2}}{U^{2}-(J')^{2}}
S_{\eta}^{\sig_{0}}\tau_{\eta}^{m_{0}}
S_{\eta+1}^{-\sig_{0}}\tau_{\eta+1}^{m_{0}}
\left.\mbox{\rule{0pt}{0.6cm}} \right\}.
\label{eqn:effec-h2}
\end{eqnarray}
\end{widetext}
The symbols $\sig_{0}=+,\,-$ and $m_{0}=+,\,-$ correspond to  the
spin $\up,\,\down$ and the orbital $A,\,B$, respectively. The
correlated hopping processes in the original system are mapped to
the spin-orbital  exchange processes in the squeezed system as
illustrated  in Fig.\ \ref{fig:hole-map}.

In the limit where $N \rightarrow \infty$ with $n$ kept constant,
the expectation values of $\nn_{i}\nn_{i+1}\delta_{ij_{\eta}}$ and
$(\nn_{i}a_{i-1}^{\da}a_{i+1}+{\rm H.c.})\delta_{ij_{\eta}}$ in
the state $|\Phi_{\rm SF}\ra$ are independent of both  $\eta$ and
$i$. Then we obtain
\begin{eqnarray}
\la H_{\rm eff}^{(2a)}\ra_{\rm SF} &=&
\frac{N}{N_{\rm e}}\la \nn_{i}\nn_{i+1}\ra_{\rm SF}
\sum_{\eta=1}^{N_{\rm e}}h(\eta,\eta+1)
\label{eqn:hole-eff2a-ab} \\
\la H_{\rm eff}^{(2b)}\ra_{\rm SF} &=&
\frac{2N}{N_{\rm e}}\la \nn_{i}a_{i-1}^{\da}a_{i+1}\ra_{\rm SF}
\sum_{\eta=1}^{N_{\rm e}}\sum_{\sig_{0}=\pm}
\sum_{m_{0}=\pm}\Theta(\eta,\sig_{0},m_{0}).
\label{eqn:effec-2b}
\end{eqnarray}
Here the notation $\la\cdot\cdot\cdot\ra_{\rm SF}$ means the
expectation value in the state $|\Phi_{\rm SF}\ra$. We note that
the matrix elements of $\Theta(\eta,\sig_{0},m_{0})$ are equal to
those of  $-h(\eta,\eta+1)/2$. We estimate the expectation values
as
\begin{eqnarray}
\la \nn_{i}\nn_{i+1} \ra_{\rm SF}
&=& n^{2}-\left(\frac{\sin n\pi}{\pi}\right)^{2}
\end{eqnarray}
and
\begin{eqnarray}
\la \nn_{i}a_{i-1}^{\da}a_{i+1}\ra_{\rm SF}
&=& n\frac{\sin 2n\pi}{2\pi}-\left(\frac{\sin n\pi}{\pi}\right)^{2},
\end{eqnarray}
by using  Wick's theorem.

Finally, we obtain the effective Hamiltonian for the spin-orbital
degrees of freedom as
\begin{eqnarray}
\la H_{\rm eff}\ra_{\rm SF}
&=&
n\left(1-\frac{\sin 2n\pi}{2n\pi}\right){\cal H}_{\rm eff}.
\label{eqn:hole-eff}
\end{eqnarray}
Here ${\cal H}_{\rm eff}$ is equal to the effective Hamiltonian
(\ref{eqn:quarter}) for the  quarter-filling, where  the summation
over lattice sites are taken over those in the squeezed system.
The factor $n(1-\sin 2n\pi/2n\pi)$ is positive for all  density
($0<n<1$).  Hence, the effective spin-orbital exchange
interactions in the hole-doped system are same with that for the
quarter-filled system except for a single multiplicative factor.
The present result suggests that the same fact  holds quite
generally for one-dimensional systems  with any number of internal
degrees of freedom.

\section{Ferromagnetism due to a single  unpaired electron in the limit
where $J-U'= \infty$}
\label{ap:J-large}

\begin{figure}
\includegraphics[width=7.5cm]{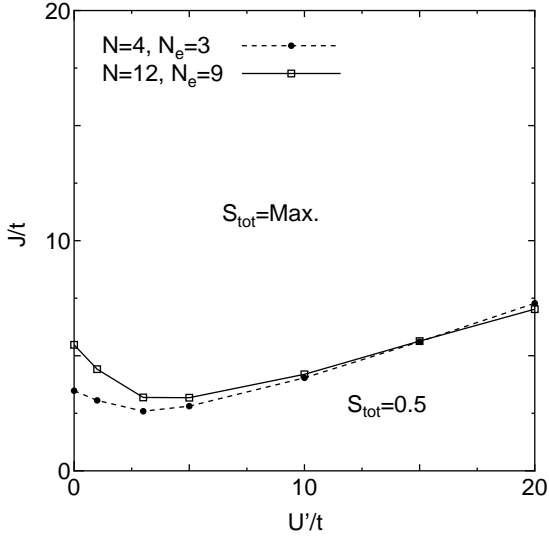}
\caption{Ground-state phase diagram for $(N,N_{\rm e})=(4,3)$ and
$(12,9)$ as examples of the odd $N_{\rm e}$ cases.}
\label{fig:Ne=odd}
\end{figure}

In finite-size calculations with odd number of electrons we find
that the ferromagnetic state is stable in a large parameter region
where $J>U'$ as shown in Fig. \ref{fig:Ne=odd}. This
ferromagnetism is caused by the motion of a single unpaired
electron among bound triplet pairs. We consider a system with odd
$N_{\rm e}$ system in the  strong coupling limit where
$J\rightarrow \infty$ with $J/U' (>1)$ kept constant under the
open boundary conditions. We prove, in the following, that the
ground state of this system is fully polarized. Only
nearest-neighbor hoppings are assumed to exist.

In this limit two electrons  on a site is bound to form a triplet
pair. As a result $M=(N_{\rm e}-1)/2$ sites are doubly occupied
and  a single unpaired electron exits.  In this limit, only
allowed processes in the Hamiltonian are  hoppings of an electron
from a doubly occupied site to a singly occupied one and those
from a singly occupied to a vacant one. The effective Hamiltonian
$H_{\rm eff}$ is written as
\begin{eqnarray}
H_{\rm eff}  =  -t\sum^{N}_{i=1}
\{P^{D}_{i}c_{im\sig}^{\da}P_{i}^{S}P^{S}_{i+1}c_{i+1m\sig}P_{i+1}^{D}
\nonumber\\
+P^{S}_{i}c_{im\sig}^{\da}P_{i}^{V}P^{V}_{i+1}c_{i+1m\sig}P_{i+1}^{S}
+\mbox{H.c.}\},
\end{eqnarray}
where $P_{i}^{D}$, $P_{i}^{S}$ and $P_{i}^{V}$ are the  projection
operators for the doubly occupied triplet, the singly-occupied
and vacant states, respectively. We can divide the whole phase
space into subspaces with fixed $N_{A}$, $N_{B}$ and $S^{z}$. Here
$N_{A}$ and $N_{B}$ denote the number of electrons in the orbital
$A$ and $B$, respectively ($|N_{A}-N_{B}|=1$), which are
conserved. Each subspace  is further divided into $_{N-1}C_{M}$
sectors according to the configuration of doubly-occupied and
vacant sites. States in each sector have the same configuration of
doubly-occupied and empty sites if we neglect the singly-occupied
site. Different sectors are disconnected with respect to $H_{\rm
eff}$. We call the sectors as  $\Gamma_{j}(N_{A},N_{B},S^{z})$,
where $j=1,2,..,_{N-1}C_{M}$. We can show  for any two
configurations $\alpha$ and $\beta$ of spins and electrons
belonging to a same sector $\Gamma_{j}(N_{A},N_{B},S^{z})$ that:
\begin{enumerate}
\item[(I)] there exists $n\ge 1$ such that
  $\la\Phi_{\alpha}|H_{\rm eff}^{n}|\Phi_{\beta}\ra\neq0 $
\item[(II)] $\la\Phi_{\alpha}|H_{\rm eff}|\Phi_{\beta}\ra\leq0. $
\end{enumerate}
Here $|\Phi_{\alpha}\ra$ denotes a basis vector chosen as
\begin{equation}
|\Phi_{\alpha}\ra =
\left[\prod_{\ell=1}^{M}(-1)^{j_{\ell}}\right]
A_{\alpha_{1}}A_{\alpha_{2}}\cdots A_{\alpha_{N}}|0\ra,
\label{eqn:baseUd=J}
\end{equation}
where $\alpha_{i}$ denotes the atomic state of the $i$-th site and
$A_{\alpha_{i}}$ is the corresponding operator, i.e.
$c_{iA\uparrow}^{\da}c_{iB\uparrow}^{\da}$,
$c_{iA\downarrow}^{\da}c_{iB\downarrow}^{\da}$,
$1/\sqrt{2}(c_{iA\uparrow}^{\da}c_{iB\downarrow}^{\da} +
c_{iA\downarrow}^{\da}c_{iB\uparrow}^{\da})$ for doubly-occupied
triplet states, $c_{im\sig}^{\da}$ for singly-occupied states, and
unity for a vacant state. The number $j_{\ell}$ denotes the
position of the $\ell$-th doubly-occupied site counted from an
end. Here $|0\ra$ is vacuum state. The proof of (I) and (II) are
straightforward\cite{Kubo}.

Now that the two conditions (I) and (II) on the matrix elements of
$H_{\rm eff}$ are satisfied, the Perron-Frobenius theorem ensures
that the ground state in the sector
$\Gamma_{j}(N_{A},N_{B},S^{z})$ is unique and can be written as a
linear combination of all the basis vectors with strictly positive
(or negative) coefficients. The positive (or negative)
definiteness of the coefficients leads to that the ground state
is an  eigenstate of the maximum value of the total
spin\cite{Kubo}.

We thus proved that the ground state in each sector is fully
polarized. As a result the ground state for subspace with fixed
$N_{A}$ and $N_{B}$ is fully polarized. We understand that the
large ferromagnetic region in the phase diagram  in the case of
odd $N_{\rm e}$ for $J- U'\gg t$ is caused by the motion of a
single unpaired electron and is a finite-size effect.


\begin{thebibliography}{99}
\bibitem{difference} In the dynamical mean-field theory, Held and
Vollhardt (HV) found ferromagnetism for the hole-doped region
($n<1$)\cite{Held} at $U=8$, $U'=6$ and $J=2$, but two of us
[Momoi and Kubo (MK)] did not discover it\cite{MomoiK} in the
region up to $U\le 40$ and $U'\le 20$. One possibility of this
discrepancy may be that ferromagnetic phase appears in a very
narrow region for the hole-doped case and MK could not discover
it. As another possibility, note that the systems studied are
different in these two cases. In ref.\ \onlinecite{Held}, HV
adopted a semi-elliptic DOS. On the other hand, in ref.\
\onlinecite{MomoiK}, MK used a Gaussian DOS, which is the exact
one for the hyper-cubic lattice in infinite dimensions. As we
showed in the present paper, the shape of DOS can greatly affect
the appearance of ferromagnetism for $n<1$. Since the long tail of
the Gaussian DOS is disadvantageous to ferromagnetism at low
densities, the discrepancy of the phase diagrams presumably comes
from the choice of DOS. Also, HV took an Ising anisotropic limit
of the Hund coupling and neglected pair-hopping terms, but MK
treated all terms correctly.
\end{thebibliography}
\end{document}